\documentclass[a4paper,11pt]{article}
\usepackage[utf8]{inputenc}
\usepackage[T1]{fontenc}
\usepackage{pos}
\usepackage{subcaption}
\usepackage{rviewport}
\usepackage{wrapfig}
\usepackage{enumitem}
\usepackage{multicol}
\usepackage{setspace}
\usepackage{xspace}
\usepackage{overpic}
\usepackage{bm}
\usepackage{booktabs}
\usepackage{siunitx}
\newcommand{\AGeVc}{\ensuremath{A\,\text{GeV}/c}\xspace}
\newcommand{\GeVc}{\ensuremath{\text{GeV}/c}\xspace}
\newcommand{\iso}[2]{\ensuremath{^{#1}\mathrm{#2}}}
\newcommand{\NASixtyOne}{NA61\slash SHINE\xspace}

\title{{Fragmentation Cross Sections for the Understanding of Cosmic-Ray Transport in the Galaxy: Results and Prospects from NA61/SHINE}}
\ShortTitle{Fragmentation Cross Sections from NA61/SHINE}

\author*[a]{Michael Unger}
\affiliation[a]{Institute for Astroparticle Physics, Karlsruhe Institute of Technology (KIT), 76131 Karlsruhe, Germany}
\onbehalf{for the NA61/SHINE Collaboration$^b$}
\affiliation[b]{{\rm\url{https://shine.web.cern.ch/node/31}}}
\author[]{}

\emailAdd{michael.unger@kit.edu}

\abstract{Accurate measurements of cosmic-ray fragmentation cross sections are essential for maximizing the physics potential of precise measurements of secondary and primary cosmic-ray fluxes from current balloon and space-borne experiments. NA61/SHINE, operating at the CERN SPS H2 beamline, is uniquely suited to studying these interactions at energies above 10 \GeVc per nucleon.

In this contribution, we present the fragmentation cross sections for the breakup of carbon into $^{10}$B, $^{11}$B and $^{11}$C at 13.5 \GeVc per nucleon that are needed for interpreting the cosmic-ray boron-to-carbon ratio. These results are based on data from a pilot run conducted in 2018.

We also give an overview of the high-statistics data-taking campaign in 2024, which covered projectile nuclei from lithium to silicon. With over 40 million recorded beam triggers, this data set will enable the reconstruction of the full reaction network required to study light secondary cosmic rays.  Furthermore, we report on data collected in 2025 with a primary oxygen beam at 150\,\GeVc per nucleon, aimed at verifying the expected flattening of fragmentation cross sections at high energies. }

\ConferenceLogo{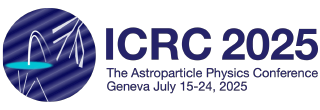}

\FullConference{39th International Cosmic Ray Conference (ICRC2025)\\
 15–24 July 2025\\
Geneva, Switzerland\\}

\begin{document}
\maketitle

\section{Introduction}
Galactic cosmic rays (GCRs) are highly relativistic nuclei, classified
as primary when directly accelerated at sources such as supernova
remnants, and secondary when produced through spallation of primaries
in the interstellar medium (ISM).  The study of secondary-to-primary
flux ratios provides important information about the propagation of
GCRs in the Galaxy.  It constrains, for example, the effective
strength and energy dependence of the diffusion coefficient for
charged particles in the Galaxy, the vertical extent of the Galactic
halo, and the expected cosmic-ray background for searches for
astrophysical dark matter.  These ratios have been measured with
percent-level precision in recent space-based experiments such as
PAMELA, AMS-02, CALET, and
DAMPE~\cite{pamelaCollab,amsCollab,caletCollab,DAMPE:2022jgy}.
However, nuclear fragmentation cross sections for the production of
secondary cosmic rays are still uncertain at the ${\sim}20\%$ level,
severely limiting the utility of these precise
measurements~\cite{genolini2018,evoli2020,Genolini:2023kcj,Maurin:2025gsz}.

Here, we present results on the cross sections for boron production in
$\iso{12}{C}+\mathrm{p}\rightarrow \mathrm{B}+X$ reactions at
13.5\,\GeVc per nucleon, obtained with the \NASixtyOne facility during
a pilot run in 2018~\cite{NA61SHINE:2024rzv}. We also describe the
full physics data set, taken in 2024 with He--Si projectiles at
12.5\,\GeVc per nucleon.

These beam momenta ($\gtrsim 10$\,\GeVc per nucleon) were chosen to address the
scarcity of precise high-energy fragmentation cross sections required
by modern GCR analyses. Furthermore, since fragmentation at these
momenta occurs in the participant--spectator (abrasion--ablation)
regime, the cross sections are expected to plateau, with only weak
energy dependence, because the underlying nucleon--nucleon inelastic
cross section varies slowly with energy.  Therefore, a single precise
measurement can strongly constrain the asymptotic energy regime. As a
cross-check of this plateau behavior, we collected additional
oxygen-fragmentation data at 150\,\GeVc per nucleon in 2025, which will also be
discussed. A concise summary of the data sets is provided in Tab.~\ref{tab:frag-overview}.
\begin{table*}[h!]
  \centering
\caption{Overview of the NA61/SHINE fragmentation runs (pilot 2018 and final physics 2024/2025).}
\label{tab:frag-overview}
  \begin{tabular}{lcccSScS}
    \toprule
    \textbf{Run} &
    \textbf{Primary} &
    \textbf{Beam} &
    \textbf{$\bm{p}_\text{beam}$} &
    {$\bm{\Sigma BL}$}&
    \textbf{$\bm{f}_\text{DAQ}$}&
    \textbf{Main} &
    \textbf{Events} \\
    &
    \textbf{beam} &
    \textbf{trigger} &
     [$\AGeVc$] &
    {[\si{Tm}]} &
    {[\si{kHz}]} &
   \textbf{trigger} &
    {[$\times10^6$]} \\
    \midrule
    2018 &  \iso{208}{Pb} & $Z_\mathrm{p}=6$ & 13.5 & 5.3 & 0.1 & beam & 0.3 \\
    2024 &  \iso{208}{Pb} & $2\leq Z_\mathrm{p} \leq 14$ & 12.5 & 5.4 & 1 & beam  & 46 \\
    2025 &  \iso{16}{O}   & $Z_\mathrm{p}=8$  & 150  & 9 & 1  &  $Z_\mathrm{f}<8$ & 5 \\
    \bottomrule
  \end{tabular}
  \vspace*{1mm}
  \parbox{0.95\textwidth}{\footnotesize
    \centering
   \scalebox{0.95}{${\Sigma BL}$: bending power; ${p}_\text{beam}$: beam momentum, $Z_\mathrm{p/f}:$ projectile/fragment charge; $f_\text{DAQ}$: readout frequency.}
  }
\end{table*}

\section[C+p interactions at 13.5\,GeV/c per nucleon]{C+p interactions at 13.5\,GeV/$\bm{c}$ per nucleon}

\subsection{Experimental Setup}
\NASixtyOne (SPS Heavy Ion and Neutrino Experiment) is located in the
CERN North Area on the H2 beam line of the Super Proton Synchrotron
(SPS)~\cite{Abgrall:2014fa}. The physics program of the experiment
includes studies of nucleus-nucleus interactions to explore phase
transitions in strongly interacting matter and investigations of
neutrino beams produced in proton-nucleus collisions. Its third main
objective concerns cosmic-ray physics, in particular hadroproduction
in air showers~\cite{NA61SHINE:2017vqs,Adhikary:2826863}, the
formation of anti-nuclei in the Galaxy~\cite{Shukla:2025imi}, and the
fragmentation of Galactic cosmic-ray nuclei, as described in these
proceedings.

\begin{figure}[t]
  \centering
  \includegraphics[clip,rviewport=0 0 0.9 1, width=\linewidth]{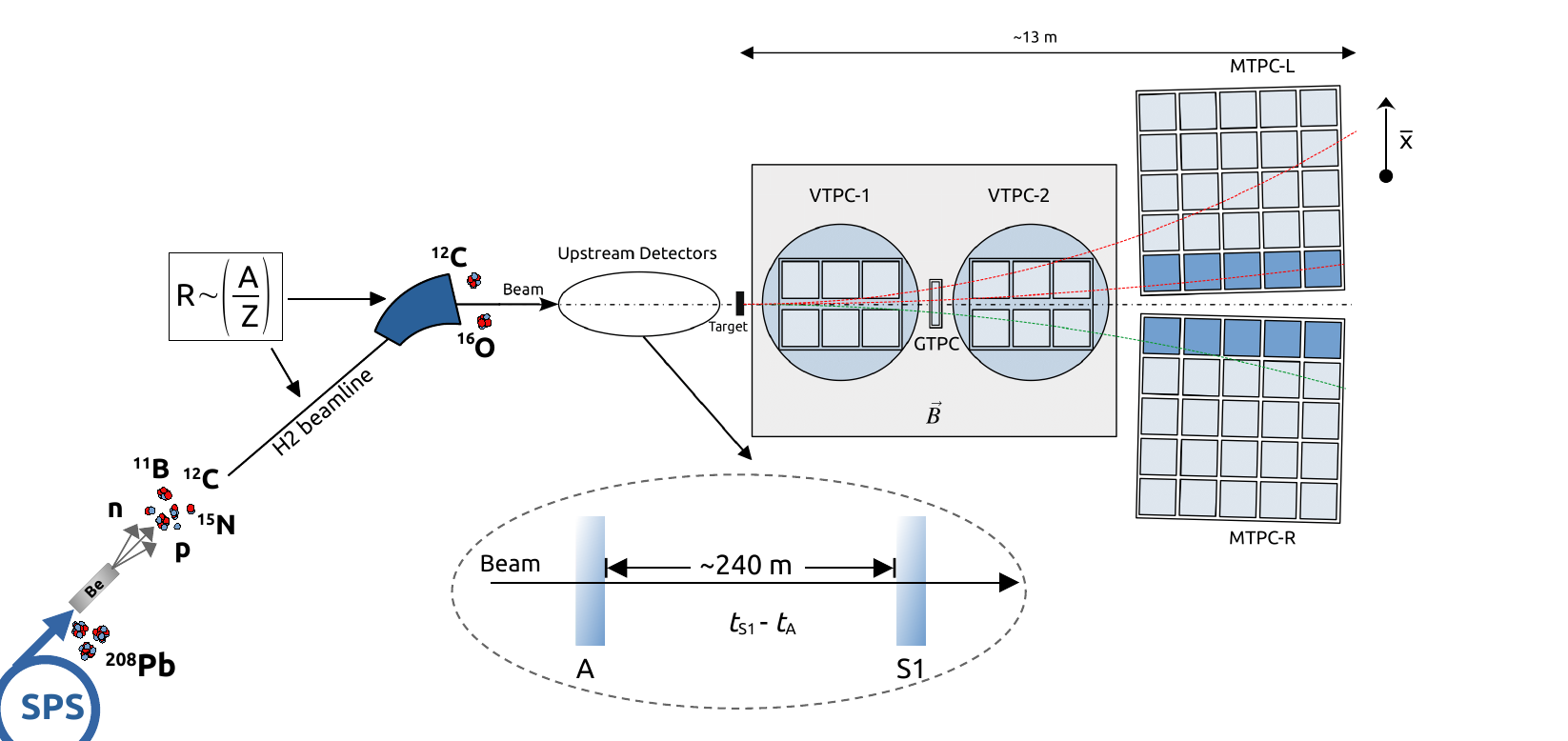}
\caption{Schematic layout of the \NASixtyOne setup at the CERN SPS, used
to measure nuclear fragmentation at 13.5\,\AGeVc during the 2018 pilot run
and at 12.5\,\AGeVc during the 2024 data taking.}
  \label{fig:setup1825}
\end{figure}

A first pilot run to study the feasibility of fragmentation measurements with
\NASixtyOne was performed in 2018. A schematic view of the experimental setup is shown in
Fig.~\ref{fig:setup1825}.

For this run, we used the secondary ion beam of the SPS. It was
produced by directing a high-intensity beam of
\iso{208}{Pb} nuclei at 13.5\,\AGeVc, extracted from the SPS, onto a
160\,mm-long beryllium plate serving as the primary target. Nuclear
fragments resulting from this \iso{208}{Pb}+Be interaction were then
transported to the \NASixtyOne experiment situated in EHN1 of the CERN
North Area, about 600\,m downstream of the T2 target, via a system of two
spectrometers that select the rigidity $R = p/Z$ of fragments with
total momentum $p$ and charge $Z$. Since the momentum per nucleon is
approximately conserved in nuclear fragmentation, the total momentum is
$p = A\,p_0/A_0$, where $A$ is the fragment mass number and $A_0$ and $p_0$
are the mass number and total momentum of the primary \iso{208}{Pb} beam,
and thus $R \varpropto A/Z$.

\begin{figure}[t]
  \centering
  \includegraphics[clip,rviewport=0 0 0.995 1, width=0.99\linewidth]{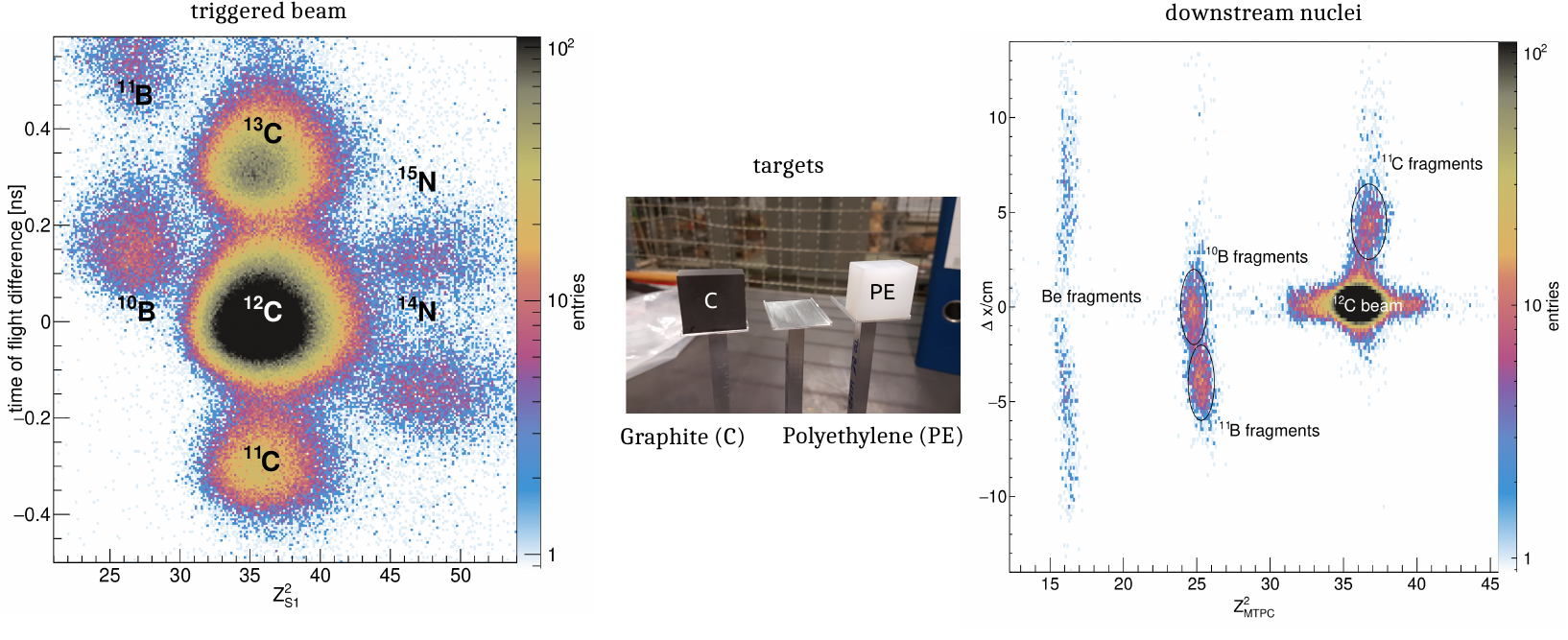}
  \caption{Particle identification upstream (left) and downstream (right) of the target (middle panel).}
  \label{fig:pid18}
\end{figure}

To study the fragmentation of \iso{12}{C}, the beam-line spectrometer
magnets were tuned to select mainly $A/Z = 2$ nuclei. The trigger was
configured to select carbon nuclei using the energy deposit in the S1
scintillator ($\propto Z^2$) upstream of the facility, while the mass
was determined via a $\sim 240$\,m time-of-flight measurement, as
indicated in the inset of Fig.~\ref{fig:setup1825}. The selection of
\iso{12}{C} beam projectiles was then performed offline, using the
distributions of these two measurements, which are shown in the left
panel of Fig.~\ref{fig:pid18}.

Since we are interested in studying the fragmentation of \iso{12}{C} on a
proton target, data taking was alternated between a polyethylene target
(hereafter abbreviated as PE) and a graphite target (hereafter abbreviated as C).
In this way, \iso{12}{C}+C interactions can be measured and subtracted
from the PE measurements. The targets are shown in the middle panel of
Fig.~\ref{fig:pid18}.

The main detectors of the facility used for this measurement consist of
three sets of Time Projection Chambers (TPCs): the Vertex TPCs
(VTPC-1/2); the Gap TPC (GTPC), located between the two VTPCs; and the
large-acceptance Main TPCs, positioned to the left and right of the
beam pipe (MTPC-L/R).

The superconducting magnets deflect the fragments with charge $Z$
produced in the beam-target interaction according to their rigidity.
Their maximum total bending power is 9\,Tm. During the 2018 pilot run, a
slightly reduced magnet current corresponding to 5.3\,Tm was used to deflect
the positively charged nuclear fragments toward the MTPC-L while
avoiding the TPC support structures. The energy deposit in the TPC gas
($\mathrm{d}E/\mathrm{d}x \propto Z^2$) and the deflection from the
nominal beam path ($\Delta x \propto Z/A$) are used for particle
identification. These measurements are shown in the right panel of
Fig.~\ref{fig:pid18}. As can be seen, the elemental separation in $Z^2$
is nearly perfect, and the isotopic composition for each element is
obtained by fitting templates to the $\Delta x$
distribution~\cite{Neeraj2025,NA61SHINE:2024rzv}.

\subsection{Charge- and Mass-Changing Cross Sections}
We present measurements of the {\itshape mass-changing} cross section,
$\upsigma_{\Delta A}$, and the {\itshape charge-changing} cross section,
$\upsigma_{\Delta Z}$, in which the projectile carbon nucleus loses at least
one nucleon or one proton, respectively~\cite{NA61SHINE:2024rzv, Marta2025,
  Neeraj2025, marta25}. The corresponding reaction is
%\begin{equation}
$
^{A}_Z\text{P} + \text{T} \rightarrow \; ^{A^{\prime}}_{Z^{\prime}}\text{P}^{\prime} + X,
$
%\end{equation}
where P denotes the projectile, T the target, and P$^\prime$ the leading
fragment in the projectile hemisphere. A mass-changing reaction has
$\Delta A = A - A^{\prime} > 0$, and a charge-changing one
$\Delta Z = Z - Z^{\prime} > 0$. P$^\prime$ is defined as the fragment with
the largest mass or charge among the projectile-side products.

\begin{figure}[t]
  \centering
  \includegraphics[width=0.99\linewidth]{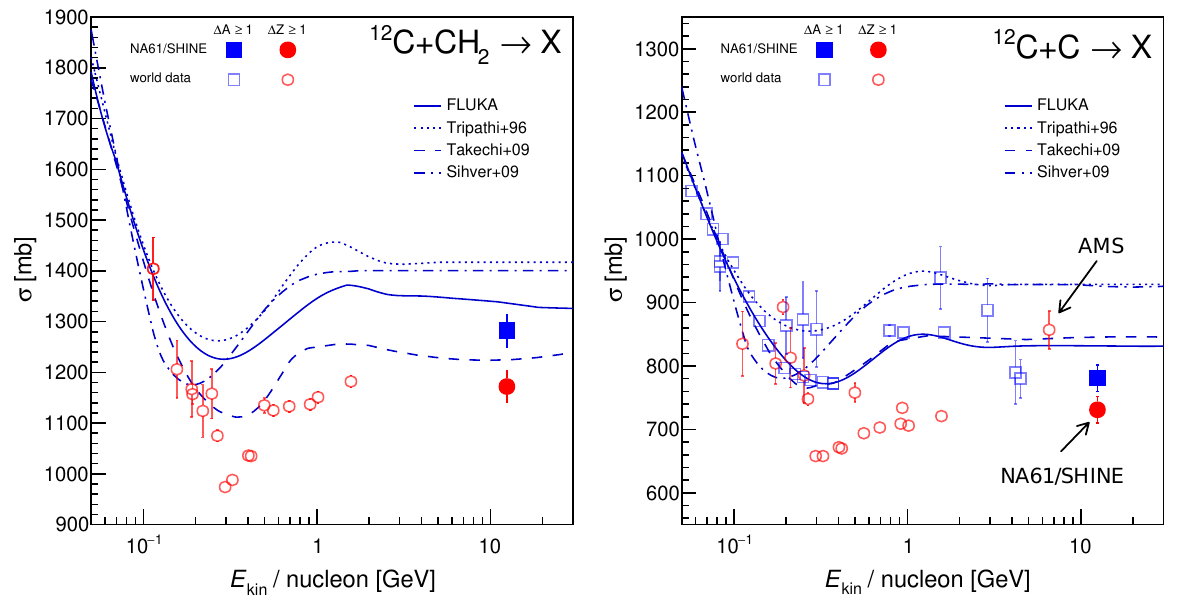}
\caption{The
   charge-changing (${\Delta{Z}\geq1}$) and mass-changing
   (${\Delta{A}\geq1}$) cross section of \iso{12}{C} on CH$_2$ (left)
   and C (right) targets as function of kinetic energy per nucleon.
   Our results \cite{NA61SHINE:2024rzv} are shown as solid circles and squares.
   Previous measurements of the charge-changing cross sections are
   shown as open circles and previous measurements of the
   mass-changing or inelastic cross section are displayed as open
   squares~\cite{Ferrando1988,Golovchenko,Webber:1990_PRC,SCHALL1996221,CHULKOV2000330,Jaros:1978,Akisenko:1980,Zhang:2002,Zheng:2002,Fang:2000,Yan:2020,Kanungo:2016}.
The charge-changing cross section derived from AMS cosmic-ray data
\cite{Yan:2020} is highlighted with an arrow. The curves are taken from
Ref.~\cite{SIHVER2012812}, showing the parametrizations of
Refs.~\cite{TRIPATHI1996347,Takechi2009,Sihver2009}, while the
FLUKA~2024.1 predictions~\cite{Ballarini:2024isa} were
taken from Ref.~\cite{flukapriv}.}
 \label{fig:ChCh}
\end{figure}

These charge- and mass-changing cross sections on C and PE targets are
compared to model predictions and previous data in
Fig.~\ref{fig:ChCh} as a function of the kinetic energy per nucleon,
$E_\text{kin} = (\sqrt{p^2 + m^2} - m)/A$, where $m$ is the nuclear
mass.

At the highest energies, the measurement derived from AMS data of the
charge-changing cross section of carbon~\cite{Yan:2020} reports $857
\pm 15\,(\text{stat.}) \pm 26\,(\text{syst.})$~mb at 7.5\,\AGeVc,
approximately constant above 4\,\AGeVc. This value significantly
exceeds our result of $731 \pm 21\,(\text{stat.}) \pm 5\,(\text{syst.})$~mb
at 13.5\,\AGeVc.

\subsection{Boron Production}
\begin{figure}[t]
  \includegraphics[width=\linewidth]{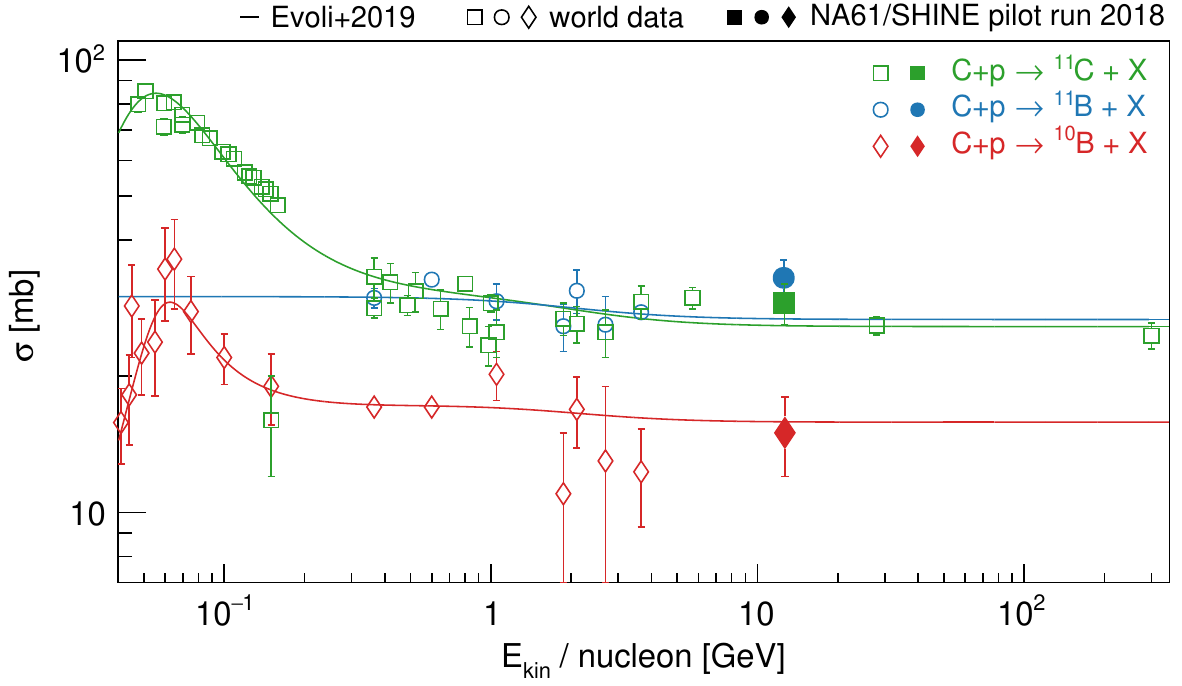}
  \caption{Isotope production cross sections of \iso{11}{C}, \iso{11}{B}
    and \iso{10}{B} fragments in \iso{12}{C}+p interactions measured with \NASixtyOne \cite{NA61SHINE:2024rzv} (solid symbols), previous
    measurements~\cite{Fontes:1977qq,Korejwo:1999,Korejwo:2002,Olson:1983,Webber:1990aipc} (open symbols) and a fit to previous data
    from Ref.~\cite{Evoli:2019pr} (lines).}
	\label{fig:bresults}
\end{figure}

The cross sections obtained from our analysis~\cite{NA61SHINE:2024rzv}
are shown in Fig.~\ref{fig:bresults} for the production of the
isotopes \iso{11}{C}, \iso{11}{B}, and \iso{10}{B} in \iso{12}{C}+p
interactions. Note that the \iso{11}{C} nucleus decays to the stable
\iso{11}{B} via $\upbeta^+$ decay with a half-life of approximately
$20$\,min, and thus contributes to the total boron
\iso{10}{B}+\iso{11}{B} production in the Galaxy.

As can be seen, previous high-energy isotopic production cross
sections for \iso{11}{C} at $28$ and
$300$\,GeV/nucleon~\cite{Cummings:1962pr, Kaufman:1976xh} are in good
agreement with our measurement. The high-energy values derived in
Ref.~\cite{Evoli:2019pr} from a fit to earlier data also agree well.
Despite being based on a short pilot run with still sizable
uncertainties, these three isotope measurements already provide meaningful
constraints on the total boron production at high energies and
demonstrate the potential of the full physics data set described in
the next section.

\section{Recent Data Sets}
\subsection[Light Nuclei at 12.5\,GeV/c per nucleon]{Light Nuclei at 12.5\,GeV/$\bm{c}$ per nucleon}
\begin{figure}[t]
  \centering
  \begin{overpic}[width=0.95\linewidth]{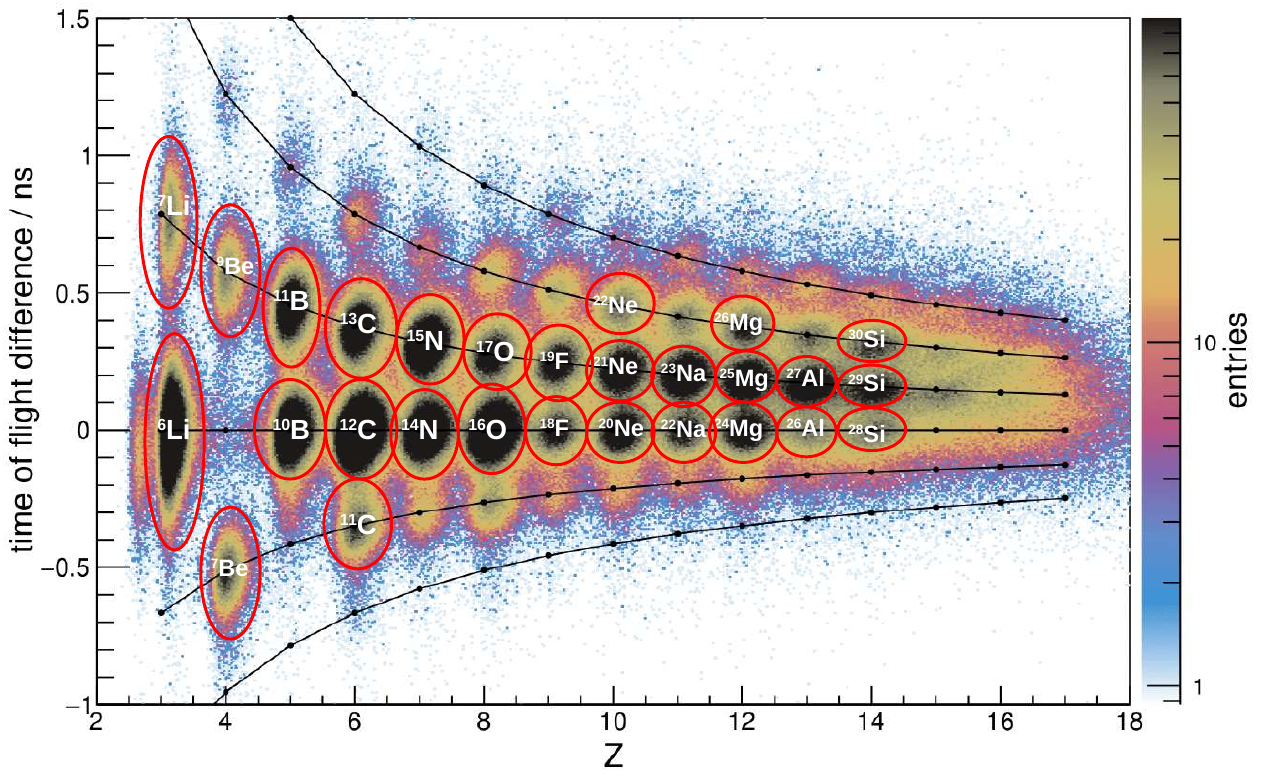}
   \put(58,57){\NASixtyOne Preliminary}
  \end{overpic}
  \caption{Beam composition during data taking in December~2024 at
12.5\,\AGeVc. The time-of-flight difference relative to fragments with
$A/Z=2$ is shown on the $y$-axis, and the charge number $Z$, measured
with a scintillator located just upstream of the NA61 target, on the
$x$-axis. Isotopes relevant for cosmic-ray
transport~\cite{Genolini:2023kcj} are labeled and highlighted by red
contours. The data shown here are based on a very preliminary calibration and
represent only a few percent of the full data set.}
  \label{fig:run24}
\end{figure}

About $46\times10^{6}$ beam triggers at 12.5~\GeVc per nucleon were recorded with
PE and C targets at the end
of 2024. The setup was very similar to the pilot run, with the
fragment tracks detected in
the MTPC-L chamber. However, data were
taken at  a more than tenfold higher rate of over 1~kHz using the
upgraded data acquisition and trigger system installed during CERN’s
Long Shutdown~2~\cite{Podlaski:2024kxg}.
 During this data taking, the
trigger was configured to record beam projectiles from He to Si ($2
\leq Z_\mathrm{p} \leq 14$). The large variety of recorded primary
isotopes is illustrated in Fig.~\ref{fig:run24}.  All isotopes relevant
for cosmic-ray transport, as discussed in Ref.~\cite{Genolini:2023kcj}, were included as projectiles.
About 10\% of the beam recorded triggers are \iso{12}{C} nuclei, i.e.\ more than an order of magnitude more events that were
used in the pilot run analysis described in the previous section. We estimate that more than $8\times10^{4}$ $\iso{12}{C}+\mathrm{p}$
interactions were collected, meeting the statistical requirements given in Ref.~\cite{Genolini:2023kcj}.

The high voltage in the sector rows of the MTPC-L was adjusted to
provide a wide dynamic range for detecting secondary particle charges
ranging from hydrogen isotopes to silicon.

\subsection[Oxygen at 150 GeV/c per nucleon]{Oxygen at 150 GeV/$\bm{c}$ per nucleon}
\begin{figure}[h]
  \def\figh{6.2cm}
\hfill \NASixtyOne Preliminary\hspace*{1cm}\\
 \includegraphics[clip,rviewport=0.15 -0.05 1 1.35,height=\figh]{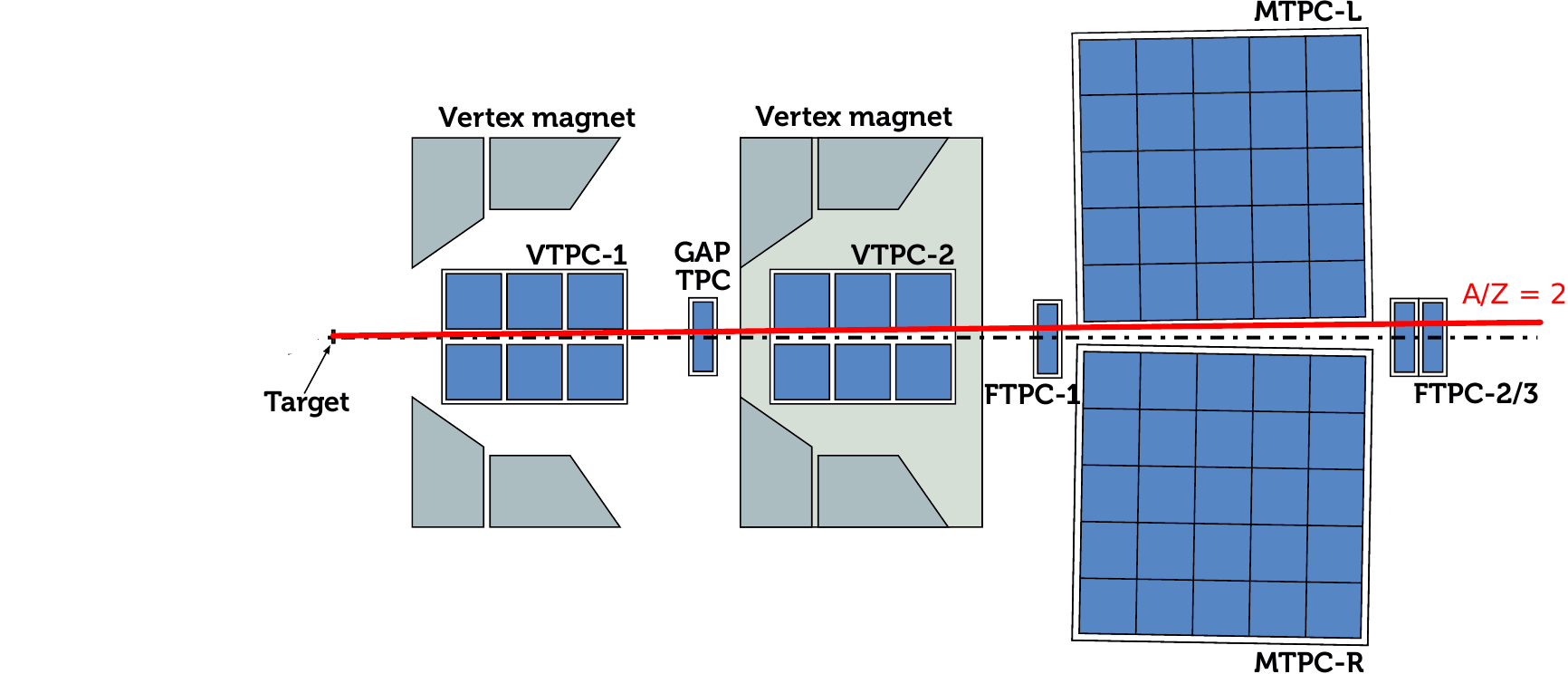}\includegraphics[height=\figh]{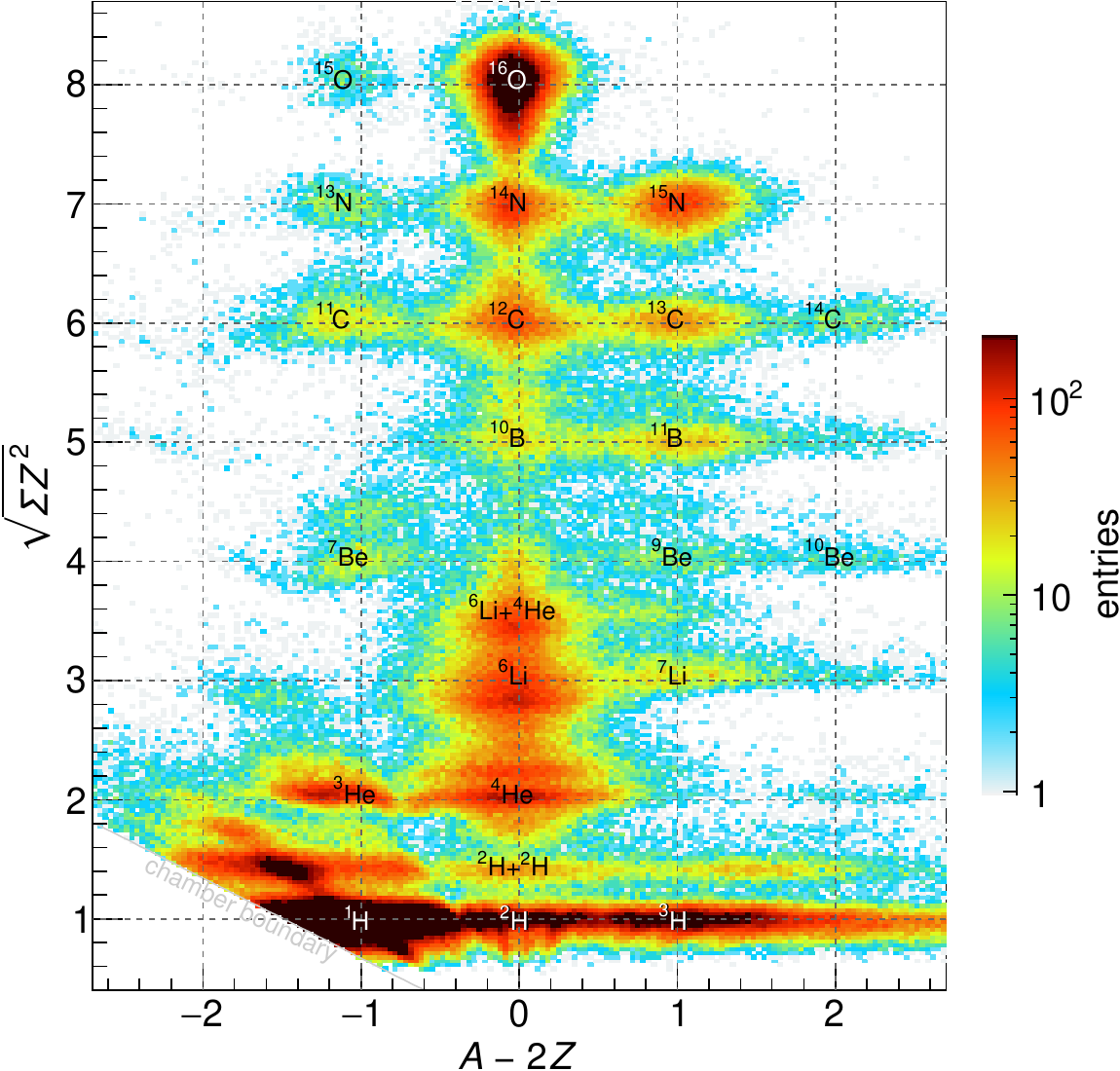}
 \caption{Left: schematic layout of the \NASixtyOne setup,
used to measure nuclear fragmentation of primary \iso{16}{O} projectiles
at 150\,\AGeVc during the 2025 run. Right: fragment charge and mass
distribution measured in the FTPC-3 chamber. The data shown are based on a
very preliminary calibration and represent less than half of the full
data set.}
  \label{fig:run25}
\end{figure}

In July 2025, we recorded $5\times10^{6}$ interactions of
\emph{primary} \iso{16}{O} at 150~\GeVc per nucleon. In this run, we employed a
charge-changing interaction trigger: the scintillator downstream of
the target was required to register a charge smaller than that of
oxygen. For this high-energy data set, the bending power of the
magnets was set to its maximum, and fragments were detected with three
Forward TPCs (FTPCs)~\cite{Rumberger:2020fwe} covering the space
between the MTPCs, as indicated in the left panel of
Fig.~\ref{fig:run25}. The resulting isotope separation is shown in the
right panel of Fig.~\ref{fig:run25}. As can be seen, even though only
one of the three chambers (FTPC-3) is used for illustration, the
isotopes are well separated. This data set will be highly valuable for
studying the energy dependence of fragmentation cross sections.

\section{Summary and Conclusions}
In these proceedings, we presented measurements of charge- and
mass-changing cross sections and isotope production in \iso{12}{C}+p
interactions at 13.5~\GeVc per nucleon from the 2018 \NASixtyOne pilot run. These
results provide new constraints on fragmentation models relevant for
Galactic cosmic-ray transport and are in good agreement with earlier
high-energy data.\\

We also reported on new measurements performed in 2024 with He--Si
projectiles at 12.5~\GeVc per nucleon and on a dedicated 2025 run with
an oxygen beam at 150~\GeVc per nucleon. The latter data set will
enable direct tests of the energy dependence of fragmentation cross
sections up to SPS energies.  Together, these measurements mark a
significant step toward a complete and precise description of the
nuclear reaction network governing secondary cosmic-ray production in
the Galaxy. The feasibility of further fragmentation measurements with
the \NASixtyOne\ setup, extending to elements up to iron after CERN’s
Long Shutdown~3, is currently being investigated.


%\cleardoublepage

\footnotesize
\setlength{\bibsep}{3pt plus 0ex}
\begin{multicols}{2}
  \setlength\columnsep{3pt}
 \setstretch{0.9}
  \bibliographystyle{uhecr}
\bibliography{main}
\end{multicols}
\section*{Acknowledgments}
We would like to thank the CERN EP, BE, HSE and EN Departments for the
strong support of NA61/SHINE.
This work was supported by
the Hungarian Scientific Research Fund (grant NKFIH 138136\slash137812\slash138152 and TKP2021-NKTA-64),
the Polish Ministry of Science and Higher Education
(DIR\slash WK\slash\-2016\slash 2017\slash\-10-1, WUT ID-UB), the National Science Centre Poland (grants
2014\slash 14\slash E\slash ST2\slash 00018, %AR, settled
2016\slash 21\slash D\slash ST2\slash 01983, %MMP, settled
2017\slash 25\slash N\slash ST2\slash 02575, %AT, settled
2018\slash 29\slash N\slash ST2\slash 02595, %AM, completed, not settled
2018\slash 30\slash A\slash ST2\slash 00226, %MG, in progress
2018\slash 31\slash G\slash ST2\slash 03910, %SK, in progress
2020\slash 39\slash O\slash ST2\slash 00277, %MR, in progress
2021\slash 43\slash P\slash ST2\slash 03319), %LT, in progress
the Norwegian Financial Mechanism 2014--2021 (grant 2019\slash 34\slash H\slash ST2\slash 00585),
the Polish Minister of Education and Science (contract No. 2021\slash WK\slash 10),
%the Russian Science Foundation (grant 17-72-20045),
%the Russian Academy of Science and the
%Russian Foundation for Basic Research (grants 08-02-00018, 09-02-00664 and 12-02-91503-CERN),
%the Russian Foundation for Basic Research (RFBR) funding within the research project no. 18-02-40086,
%the Ministry of Science and Higher Education of the Russian Federation, Project "Fundamental properties of elementary particles and cosmology" No 0723-2020-0041,
the European Union's Horizon 2020 research and innovation programme under grant agreement No. 871072,
the Ministry of Education, Culture, Sports,
Science and Tech\-no\-lo\-gy, Japan, Grant-in-Aid for Sci\-en\-ti\-fic
Research (grants 18071005, 19034011, 19740162, 20740160 and 20039012,22H04943),
the German Research Foundation DFG (grants GA\,1480\slash8-1 and project 426579465),
the Bulgarian Ministry of Education and Science within the National
Roadmap for Research Infrastructures 2020--2027, contract No. D01-374/18.12.2020,
Serbian Ministry of Science, Technological Development and Innovation (grant
OI171002), Swiss Nationalfonds Foundation (grant 200020\-117913/1),
ETH Research Grant TH-01\,07-3, National Science Foundation grant
PHY-2013228 and the Fermi National Accelerator Laboratory (Fermilab),
a U.S. Department of Energy, Office of Science, HEP User Facility
managed by Fermi Research Alliance, LLC (FRA), acting under Contract
No. DE-AC02-07CH11359 and the IN2P3-CNRS (France).\\

% The data used in this paper were collected before February 2022.

\section*{The \NASixtyOne Collaboration}
\bigskip
\scriptsize
\begin{sloppypar}
% based on XML DB with time Mon Oct 13 18:41:50 2025
% Authors in alphabetical order.
\noindent
{H.\;Adhikary~\href{https://orcid.org/0000-0002-5746-1268}{\includegraphics[height=1.7ex]{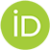}}\textsuperscript{\,11}},
{P.\;Adrich~\href{https://orcid.org/0000-0002-7019-5451}{\includegraphics[height=1.7ex]{orcid-logo.png}}\textsuperscript{\,13}},
{K.K.\;Allison~\href{https://orcid.org/0000-0002-3494-9383}{\includegraphics[height=1.7ex]{orcid-logo.png}}\textsuperscript{\,24}},
{N.\;Amin~\href{https://orcid.org/0009-0004-7572-3817}{\includegraphics[height=1.7ex]{orcid-logo.png}}\textsuperscript{\,4}},
{E.V.\;Andronov~\href{https://orcid.org/0000-0003-0437-9292}{\includegraphics[height=1.7ex]{orcid-logo.png}}\textsuperscript{\,21}},
{I.-C.\;Arsene~\href{https://orcid.org/0000-0003-2316-9565}{\includegraphics[height=1.7ex]{orcid-logo.png}}\textsuperscript{\,10}},
{M.\;Bajda~\href{https://orcid.org/0009-0005-8859-1099}{\includegraphics[height=1.7ex]{orcid-logo.png}}\textsuperscript{\,14}},
{Y.\;Balkova~\href{https://orcid.org/0000-0002-6957-573X}{\includegraphics[height=1.7ex]{orcid-logo.png}}\textsuperscript{\,13}},
{D.\;Battaglia~\href{https://orcid.org/0000-0002-5283-0992}{\includegraphics[height=1.7ex]{orcid-logo.png}}\textsuperscript{\,23}},
{A.\;Bazgir~\href{https://orcid.org/0000-0003-0358-0576}{\includegraphics[height=1.7ex]{orcid-logo.png}}\textsuperscript{\,11}},
{S.\;Bhosale~\href{https://orcid.org/0000-0001-5709-4747}{\includegraphics[height=1.7ex]{orcid-logo.png}}\textsuperscript{\,12}},
{M.\;Bielewicz~\href{https://orcid.org/0000-0001-8267-4874}{\includegraphics[height=1.7ex]{orcid-logo.png}}\textsuperscript{\,13}},
{A.\;Blondel~\href{https://orcid.org/0000-0002-1597-8859}{\includegraphics[height=1.7ex]{orcid-logo.png}}\textsuperscript{\,3}},
{M.\;Bogomilov~\href{https://orcid.org/0000-0001-7738-2041}{\includegraphics[height=1.7ex]{orcid-logo.png}}\textsuperscript{\,2}},
{Y.\;Bondar~\href{https://orcid.org/0000-0003-2773-9668}{\includegraphics[height=1.7ex]{orcid-logo.png}}\textsuperscript{\,11}},
{J.\;Brzychczyk~\href{https://orcid.org/0000-0001-5320-6748}{\includegraphics[height=1.7ex]{orcid-logo.png}}\textsuperscript{\,14}},
{M.\;Buryakov~\href{https://orcid.org/0009-0008-2394-4967}{\includegraphics[height=1.7ex]{orcid-logo.png}}\textsuperscript{\,20}},
{A.F.\;Camino\textsuperscript{\,26}},
{Y.D.\;Chandak~\href{https://orcid.org/0009-0009-2080-566X}{\includegraphics[height=1.7ex]{orcid-logo.png}}\textsuperscript{\,24}},
{M.\;Csan\'ad~\href{https://orcid.org/0000-0002-3154-6925}{\includegraphics[height=1.7ex]{orcid-logo.png}}\textsuperscript{\,6}},
{M.\;\'Cwiok~\href{https://orcid.org/0009-0007-1440-9113}{\includegraphics[height=1.7ex]{orcid-logo.png}}\textsuperscript{\,17}},
{T.\;Czopowicz~\href{https://orcid.org/0000-0003-1908-2977}{\includegraphics[height=1.7ex]{orcid-logo.png}}\textsuperscript{\,13}},
{C.\;Dalmazzone~\href{https://orcid.org/0000-0001-6945-5845}{\includegraphics[height=1.7ex]{orcid-logo.png}}\textsuperscript{\,3}},
{N.\;Davis~\href{https://orcid.org/0000-0003-3047-6854}{\includegraphics[height=1.7ex]{orcid-logo.png}}\textsuperscript{\,12}},
{A.\;Dmitriev~\href{https://orcid.org/0000-0001-7853-0173}{\includegraphics[height=1.7ex]{orcid-logo.png}}\textsuperscript{\,20}},
{P.~von\;Doetinchem~\href{https://orcid.org/0000-0002-7801-3376}{\includegraphics[height=1.7ex]{orcid-logo.png}}\textsuperscript{\,25}},
{W.\;Dominik~\href{https://orcid.org/0000-0001-7444-9239}{\includegraphics[height=1.7ex]{orcid-logo.png}}\textsuperscript{\,17}},
{J.\;Dumarchez~\href{https://orcid.org/0000-0002-9243-4425}{\includegraphics[height=1.7ex]{orcid-logo.png}}\textsuperscript{\,3}},
{R.\;Engel~\href{https://orcid.org/0000-0003-2924-8889}{\includegraphics[height=1.7ex]{orcid-logo.png}}\textsuperscript{\,4}},
{G.A.\;Feofilov~\href{https://orcid.org/0000-0003-3700-8623}{\includegraphics[height=1.7ex]{orcid-logo.png}}\textsuperscript{\,21}},
{L.\;Fields~\href{https://orcid.org/0000-0001-8281-3686}{\includegraphics[height=1.7ex]{orcid-logo.png}}\textsuperscript{\,23}},
{Z.\;Fodor~\href{https://orcid.org/0000-0003-2519-5687}{\includegraphics[height=1.7ex]{orcid-logo.png}}\textsuperscript{\,5,18}},
{M.\;Friend~\href{https://orcid.org/0000-0003-4660-4670}{\includegraphics[height=1.7ex]{orcid-logo.png}}\textsuperscript{\,7}},
{M.\;Ga\'zdzicki~\href{https://orcid.org/0000-0002-6114-8223}{\includegraphics[height=1.7ex]{orcid-logo.png}}\textsuperscript{\,11}},
{K.E.\;Gollwitzer\textsuperscript{\,22}},
{O.\;Golosov~\href{https://orcid.org/0000-0001-6562-2925}{\includegraphics[height=1.7ex]{orcid-logo.png}}\textsuperscript{\,21}},
{V.\;Golovatyuk~\href{https://orcid.org/0009-0006-5201-0990}{\includegraphics[height=1.7ex]{orcid-logo.png}}\textsuperscript{\,20}},
{M.\;Golubeva~\href{https://orcid.org/0009-0003-4756-2449}{\includegraphics[height=1.7ex]{orcid-logo.png}}\textsuperscript{\,21}},
{K.\;Grebieszkow~\href{https://orcid.org/0000-0002-6754-9554}{\includegraphics[height=1.7ex]{orcid-logo.png}}\textsuperscript{\,19}},
{F.\;Guber~\href{https://orcid.org/0000-0001-8790-3218}{\includegraphics[height=1.7ex]{orcid-logo.png}}\textsuperscript{\,21}},
{P.G.\;Hurh~\href{https://orcid.org/0000-0002-9024-5399}{\includegraphics[height=1.7ex]{orcid-logo.png}}\textsuperscript{\,22}},
{S.\;Ilieva~\href{https://orcid.org/0000-0001-9204-2563}{\includegraphics[height=1.7ex]{orcid-logo.png}}\textsuperscript{\,2}},
{A.\;Ivashkin~\href{https://orcid.org/0000-0003-4595-5866}{\includegraphics[height=1.7ex]{orcid-logo.png}}\textsuperscript{\,21}},
{N.\;Karpushkin~\href{https://orcid.org/0000-0001-5513-9331}{\includegraphics[height=1.7ex]{orcid-logo.png}}\textsuperscript{\,21}},
{M.\;Kie{\l}bowicz~\href{https://orcid.org/0000-0002-4403-9201}{\includegraphics[height=1.7ex]{orcid-logo.png}}\textsuperscript{\,12}},
{V.A.\;Kireyeu~\href{https://orcid.org/0000-0002-5630-9264}{\includegraphics[height=1.7ex]{orcid-logo.png}}\textsuperscript{\,20}},
{R.\;Kolesnikov~\href{https://orcid.org/0009-0006-4224-1058}{\includegraphics[height=1.7ex]{orcid-logo.png}}\textsuperscript{\,20}},
{D.\;Kolev~\href{https://orcid.org/0000-0002-9203-4739}{\includegraphics[height=1.7ex]{orcid-logo.png}}\textsuperscript{\,2}},
{Y.\;Koshio~\href{https://orcid.org/0000-0003-0437-8505}{\includegraphics[height=1.7ex]{orcid-logo.png}}\textsuperscript{\,8}},
{S.\;Kowalski~\href{https://orcid.org/0000-0001-9888-4008}{\includegraphics[height=1.7ex]{orcid-logo.png}}\textsuperscript{\,16}},
{T.\;Kowalski~\href{https://orcid.org/0000-0002-2550-1704}{\includegraphics[height=1.7ex]{orcid-logo.png}}\textsuperscript{\,13}},
{B.\;Koz{\l}owski~\href{https://orcid.org/0000-0001-8442-2320}{\includegraphics[height=1.7ex]{orcid-logo.png}}\textsuperscript{\,19}},
{A.\;Krasnoperov~\href{https://orcid.org/0000-0002-1425-2861}{\includegraphics[height=1.7ex]{orcid-logo.png}}\textsuperscript{\,20}},
{W.\;Kucewicz~\href{https://orcid.org/0000-0002-2073-711X}{\includegraphics[height=1.7ex]{orcid-logo.png}}\textsuperscript{\,15}},
{M.\;Kuchowicz~\href{https://orcid.org/0000-0003-3174-585X}{\includegraphics[height=1.7ex]{orcid-logo.png}}\textsuperscript{\,18}},
{P.\;Lasko~\href{https://orcid.org/0000-0003-1110-9522}{\includegraphics[height=1.7ex]{orcid-logo.png}}\textsuperscript{\,14}},
{A.\;L\'aszl\'o~\href{https://orcid.org/0000-0003-2712-6968}{\includegraphics[height=1.7ex]{orcid-logo.png}}\textsuperscript{\,5}},
{M.\;Lewicki~\href{https://orcid.org/0000-0002-8972-3066}{\includegraphics[height=1.7ex]{orcid-logo.png}}\textsuperscript{\,12}},
{G.\;Lykasov~\href{https://orcid.org/0000-0002-1544-6959}{\includegraphics[height=1.7ex]{orcid-logo.png}}\textsuperscript{\,20}},
{J.R.\;Lyon~\href{https://orcid.org/0009-0003-2579-8821}{\includegraphics[height=1.7ex]{orcid-logo.png}}\textsuperscript{\,25}},
{V.V.\;Lyubushkin~\href{https://orcid.org/0000-0003-0136-233X}{\includegraphics[height=1.7ex]{orcid-logo.png}}\textsuperscript{\,20}},
{M.\;Ma\'ckowiak-Paw{\l}owska~\href{https://orcid.org/0000-0003-3954-6329}{\includegraphics[height=1.7ex]{orcid-logo.png}}\textsuperscript{\,19}},
{B.\;Maksiak~\href{https://orcid.org/0000-0002-7950-2307}{\includegraphics[height=1.7ex]{orcid-logo.png}}\textsuperscript{\,13}},
{A.I.\;Malakhov~\href{https://orcid.org/0000-0001-8569-8409}{\includegraphics[height=1.7ex]{orcid-logo.png}}\textsuperscript{\,20}},
{A.\;Marcinek~\href{https://orcid.org/0000-0001-9922-743X}{\includegraphics[height=1.7ex]{orcid-logo.png}}\textsuperscript{\,12}},
{A.D.\;Marino~\href{https://orcid.org/0000-0002-1709-538X}{\includegraphics[height=1.7ex]{orcid-logo.png}}\textsuperscript{\,24}},
{T.\;Matulewicz~\href{https://orcid.org/0000-0003-2098-1216}{\includegraphics[height=1.7ex]{orcid-logo.png}}\textsuperscript{\,17}},
{V.\;Matveev~\href{https://orcid.org/0000-0002-2745-5908}{\includegraphics[height=1.7ex]{orcid-logo.png}}\textsuperscript{\,20}},
{G.L.\;Melkumov~\href{https://orcid.org/0009-0004-2074-6755}{\includegraphics[height=1.7ex]{orcid-logo.png}}\textsuperscript{\,20}},
{A.\;Merzlaya~\href{https://orcid.org/0000-0002-6553-2783}{\includegraphics[height=1.7ex]{orcid-logo.png}}\textsuperscript{\,10}},
{{\L}.\;Mik~\href{https://orcid.org/0000-0003-2712-6861}{\includegraphics[height=1.7ex]{orcid-logo.png}}\textsuperscript{\,15}},
{S.\;Morozov~\href{https://orcid.org/0000-0002-6748-7277}{\includegraphics[height=1.7ex]{orcid-logo.png}}\textsuperscript{\,21}},
{Y.\;Nagai~\href{https://orcid.org/0000-0002-1792-5005}{\includegraphics[height=1.7ex]{orcid-logo.png}}\textsuperscript{\,6}},
{R.\;Nagy~\href{https://orcid.org/0009-0004-4274-1832}{\includegraphics[height=1.7ex]{orcid-logo.png}}\textsuperscript{\,5}},
{T.\;Nakadaira~\href{https://orcid.org/0000-0003-4327-7598}{\includegraphics[height=1.7ex]{orcid-logo.png}}\textsuperscript{\,7}},
{S.\;Nishimori~\href{https://orcid.org/~0000-0002-1820-0938}{\includegraphics[height=1.7ex]{orcid-logo.png}}\textsuperscript{\,7}},
{A.\;Olivier~\href{https://orcid.org/0000-0003-4261-8303}{\includegraphics[height=1.7ex]{orcid-logo.png}}\textsuperscript{\,23}},
{V.\;Ozvenchuk~\href{https://orcid.org/0000-0002-7821-7109}{\includegraphics[height=1.7ex]{orcid-logo.png}}\textsuperscript{\,12}},
{O.\;Panova~\href{https://orcid.org/0000-0001-5039-7788}{\includegraphics[height=1.7ex]{orcid-logo.png}}\textsuperscript{\,11}},
{V.\;Paolone~\href{https://orcid.org/0000-0003-2162-0957}{\includegraphics[height=1.7ex]{orcid-logo.png}}\textsuperscript{\,26}},
{I.\;Pidhurskyi~\href{https://orcid.org/0000-0001-9916-9436}{\includegraphics[height=1.7ex]{orcid-logo.png}}\textsuperscript{\,11}},
{R.\;P{\l}aneta~\href{https://orcid.org/0000-0001-8007-8577}{\includegraphics[height=1.7ex]{orcid-logo.png}}\textsuperscript{\,14}},
{P.\;Podlaski~\href{https://orcid.org/0000-0002-0232-9841}{\includegraphics[height=1.7ex]{orcid-logo.png}}\textsuperscript{\,17}},
{B.A.\;Popov~\href{https://orcid.org/0000-0001-5416-9301}{\includegraphics[height=1.7ex]{orcid-logo.png}}\textsuperscript{\,20,3}},
{B.\;P\'orfy~\href{https://orcid.org/0000-0001-5724-9737}{\includegraphics[height=1.7ex]{orcid-logo.png}}\textsuperscript{\,5,6}},
{D.S.\;Prokhorova~\href{https://orcid.org/0000-0003-3726-9196}{\includegraphics[height=1.7ex]{orcid-logo.png}}\textsuperscript{\,21}},
{D.\;Pszczel~\href{https://orcid.org/0000-0002-4697-6688}{\includegraphics[height=1.7ex]{orcid-logo.png}}\textsuperscript{\,13}},
{S.\;Pu{\l}awski~\href{https://orcid.org/0000-0003-1982-2787}{\includegraphics[height=1.7ex]{orcid-logo.png}}\textsuperscript{\,16}},
{L.\;Ren~\href{https://orcid.org/0000-0003-1709-7673}{\includegraphics[height=1.7ex]{orcid-logo.png}}\textsuperscript{\,24}},
{V.Z.\;Reyna~Ortiz~\href{https://orcid.org/0000-0002-7026-8198}{\includegraphics[height=1.7ex]{orcid-logo.png}}\textsuperscript{\,11}},
{D.\;R\"ohrich\textsuperscript{\,9}},
{M.\;Roth~\href{https://orcid.org/0000-0003-1281-4477}{\includegraphics[height=1.7ex]{orcid-logo.png}}\textsuperscript{\,4}},
{{\L}.\;Rozp{\l}ochowski~\href{https://orcid.org/0000-0003-3680-6738}{\includegraphics[height=1.7ex]{orcid-logo.png}}\textsuperscript{\,12}},
{M.\;Rumyantsev~\href{https://orcid.org/0000-0001-8233-2030}{\includegraphics[height=1.7ex]{orcid-logo.png}}\textsuperscript{\,20}},
{A.\;Rustamov~\href{https://orcid.org/0000-0001-8678-6400}{\includegraphics[height=1.7ex]{orcid-logo.png}}\textsuperscript{\,1}},
{M.\;Rybczynski~\href{https://orcid.org/0000-0002-3638-3766}{\includegraphics[height=1.7ex]{orcid-logo.png}}\textsuperscript{\,11}},
{A.\;Rybicki~\href{https://orcid.org/0000-0003-3076-0505}{\includegraphics[height=1.7ex]{orcid-logo.png}}\textsuperscript{\,12}},
{D.\;Rybka~\href{https://orcid.org/0000-0002-9924-6398}{\includegraphics[height=1.7ex]{orcid-logo.png}}\textsuperscript{\,13}},
{K.\;Sakashita~\href{https://orcid.org/0000-0003-2602-7837}{\includegraphics[height=1.7ex]{orcid-logo.png}}\textsuperscript{\,7}},
{K.\;Schmidt~\href{https://orcid.org/0000-0002-0903-5790}{\includegraphics[height=1.7ex]{orcid-logo.png}}\textsuperscript{\,16}},
{P.\;Seyboth~\href{https://orcid.org/0000-0002-4821-6105}{\includegraphics[height=1.7ex]{orcid-logo.png}}\textsuperscript{\,11}},
{U.A.\;Shah~\href{https://orcid.org/0000-0002-9315-1304}{\includegraphics[height=1.7ex]{orcid-logo.png}}\textsuperscript{\,11}},
{Y.\;Shiraishi~\href{https://orcid.org/0000-0002-0132-3923}{\includegraphics[height=1.7ex]{orcid-logo.png}}\textsuperscript{\,8}},
{A.\;Shukla~\href{https://orcid.org/0000-0003-3839-7229}{\includegraphics[height=1.7ex]{orcid-logo.png}}\textsuperscript{\,25}},
{M.\;S{\l}odkowski~\href{https://orcid.org/0000-0003-0463-2753}{\includegraphics[height=1.7ex]{orcid-logo.png}}\textsuperscript{\,19}},
{P.\;Staszel~\href{https://orcid.org/0000-0003-4002-1626}{\includegraphics[height=1.7ex]{orcid-logo.png}}\textsuperscript{\,14}},
{G.\;Stefanek~\href{https://orcid.org/0000-0001-6656-9177}{\includegraphics[height=1.7ex]{orcid-logo.png}}\textsuperscript{\,11}},
{J.\;Stepaniak~\href{https://orcid.org/0000-0003-2064-9870}{\includegraphics[height=1.7ex]{orcid-logo.png}}\textsuperscript{\,13}},
{{\L}.\;\'Swiderski~\href{https://orcid.org/0000-0001-5857-2085}{\includegraphics[height=1.7ex]{orcid-logo.png}}\textsuperscript{\,13}},
{J.\;Szewi\'nski~\href{https://orcid.org/0000-0003-2981-9303}{\includegraphics[height=1.7ex]{orcid-logo.png}}\textsuperscript{\,13}},
{R.\;Szukiewicz~\href{https://orcid.org/0000-0002-1291-4040}{\includegraphics[height=1.7ex]{orcid-logo.png}}\textsuperscript{\,18}},
{A.\;Taranenko~\href{https://orcid.org/0000-0003-1737-4474}{\includegraphics[height=1.7ex]{orcid-logo.png}}\textsuperscript{\,21}},
{A.\;Tefelska~\href{https://orcid.org/0000-0002-6069-4273}{\includegraphics[height=1.7ex]{orcid-logo.png}}\textsuperscript{\,19}},
{D.\;Tefelski~\href{https://orcid.org/0000-0003-0802-2290}{\includegraphics[height=1.7ex]{orcid-logo.png}}\textsuperscript{\,19}},
{V.\;Tereshchenko~\href{https://orcid.org/0000-0001-8996-2254}{\includegraphics[height=1.7ex]{orcid-logo.png}}\textsuperscript{\,20}},
{R.\;Tsenov~\href{https://orcid.org/0000-0002-1330-8640}{\includegraphics[height=1.7ex]{orcid-logo.png}}\textsuperscript{\,2}},
{L.\;Turko~\href{https://orcid.org/0000-0002-5474-8650}{\includegraphics[height=1.7ex]{orcid-logo.png}}\textsuperscript{\,18}},
{T.S.\;Tveter~\href{https://orcid.org/0009-0003-7140-8644}{\includegraphics[height=1.7ex]{orcid-logo.png}}\textsuperscript{\,10}},
{M.\;Unger~\href{https://orcid.org/0000-0002-7651-0272}{\includegraphics[height=1.7ex]{orcid-logo.png}}\textsuperscript{\,4}},
{M.\;Urbaniak~\href{https://orcid.org/0000-0002-9768-030X}{\includegraphics[height=1.7ex]{orcid-logo.png}}\textsuperscript{\,16}},
{D.\;Veberi\v{c}~\href{https://orcid.org/0000-0003-2683-1526}{\includegraphics[height=1.7ex]{orcid-logo.png}}\textsuperscript{\,4}},
{O.\;Vitiuk~\href{https://orcid.org/0000-0002-9744-3937}{\includegraphics[height=1.7ex]{orcid-logo.png}}\textsuperscript{\,18}},
{A.\;Wickremasinghe~\href{https://orcid.org/0000-0002-5325-0455}{\includegraphics[height=1.7ex]{orcid-logo.png}}\textsuperscript{\,22}},
{K.\;Witek~\href{https://orcid.org/0009-0004-6699-1895}{\includegraphics[height=1.7ex]{orcid-logo.png}}\textsuperscript{\,15}},
{K.\;W\'ojcik~\href{https://orcid.org/0000-0002-8315-9281}{\includegraphics[height=1.7ex]{orcid-logo.png}}\textsuperscript{\,16}},
{O.\;Wyszy\'nski~\href{https://orcid.org/0000-0002-6652-0450}{\includegraphics[height=1.7ex]{orcid-logo.png}}\textsuperscript{\,11}},
{A.\;Zaitsev~\href{https://orcid.org/0000-0003-4711-9925}{\includegraphics[height=1.7ex]{orcid-logo.png}}\textsuperscript{\,20}},
{E.\;Zherebtsova~\href{https://orcid.org/0000-0002-1364-0969}{\includegraphics[height=1.7ex]{orcid-logo.png}}\textsuperscript{\,18}},
{E.D.\;Zimmerman~\href{https://orcid.org/0000-0002-6394-6659}{\includegraphics[height=1.7ex]{orcid-logo.png}}\textsuperscript{\,24}}, and
{A.\;Zviagina~\href{https://orcid.org/0009-0007-5211-6493}{\includegraphics[height=1.7ex]{orcid-logo.png}}\textsuperscript{\,21}}

\end{sloppypar}
\begin{multicols}{2}
  \setlength\columnsep{3pt}
 \setstretch{0.9}
% based on XML DB with time Mon Oct 13 18:41:50 2025
% Institutes in alphabetical order.

\noindent
\textsuperscript{1}~National Nuclear Research Center, Baku, Azerbaijan\\
\textsuperscript{2}~Faculty of Physics, University of Sofia, Sofia, Bulgaria\\
\textsuperscript{3}~LPNHE, Sorbonne University, CNRS/IN2P3, Paris, France\\
\textsuperscript{4}~Karlsruhe Institute of Technology, Karlsruhe, Germany\\
\textsuperscript{5}~HUN-REN Wigner Research Centre for Physics, Budapest, Hungary\\
\textsuperscript{6}~E\"otv\"os Lor\'and University, Budapest, Hungary\\
\textsuperscript{7}~Institute for Particle and Nuclear Studies, Tsukuba, Japan\\
\textsuperscript{8}~Okayama University, Okayama, Japan\\
\textsuperscript{9}~University of Bergen, Bergen, Norway\\
\textsuperscript{10}~University of Oslo, Oslo, Norway\\
\textsuperscript{11}~Jan Kochanowski University, Kielce, Poland\\
\textsuperscript{12}~Institute of Nuclear Physics, Polish Academy of Sciences, Cracow, Poland\\
\textsuperscript{13}~National Centre for Nuclear Research, Warsaw, Poland\\
\textsuperscript{14}~Jagiellonian University, Cracow, Poland\\
\textsuperscript{15}~AGH - University of Krakow, Cracow, Poland\\
\textsuperscript{16}~University of Silesia, Katowice, Poland\\
\textsuperscript{17}~University of Warsaw, Warsaw, Poland\\
\textsuperscript{18}~University of Wroc{\l}aw,  Wroc{\l}aw, Poland\\
\textsuperscript{19}~Warsaw University of Technology, Warsaw, Poland\\
\textsuperscript{20}~Joint Institute for Nuclear Research, Dubna, International Organization\\
\textsuperscript{21}~Affiliated with an institution formerly covered by a cooperation agreement with CERN\\
\textsuperscript{22}~Fermilab, Batavia, USA\\
\textsuperscript{23}~University of Notre Dame, Notre Dame, USA\\
\textsuperscript{24}~University of Colorado, Boulder, USA\\
\textsuperscript{25}~University of Hawaii at Manoa, Honolulu, USA\\
\textsuperscript{26}~University of Pittsburgh, Pittsburgh, USA\\

\end{multicols}

\end{document}